\documentclass[aps,pra,twocolumn,superscriptaddress]{revtex4-1}

\usepackage{graphicx}
\usepackage{dcolumn}
\usepackage{bm}
\usepackage{amssymb}
\usepackage{amsmath}
\usepackage{subfigure}

\usepackage{url}
\usepackage[breaklinks]{hyperref}
\usepackage{breakurl}
\usepackage{color}
\newif\ifhyper
\hypertrue
\ifhyper
\hypersetup{
   citecolor = {green},
   colorlinks = {true},
   urlcolor = {blue}
}
\fi

\usepackage{color}
\usepackage{soul}

\begin{document}

\title{Towards Prediction of Financial Crashes with a D-Wave Quantum Annealer }

\author{Yongcheng Ding}
\affiliation{International Center of Quantum Artificial Intelligence for Science and Technology (QuArtist) \\ and Department of Physics, Shanghai University, 200444 Shanghai, China}
\affiliation{Department of Physical Chemistry, University of the Basque Country UPV/EHU, Apartado 644, 48080 Bilbao, Spain}
\affiliation{ProQuam Co., Ltd., 200444 Shanghai, China.}
\author{Javier Gonzalez-Conde}
\email{javier.gonzalezc@ehu.eus}
\affiliation{Department of Physical Chemistry, University of the Basque Country UPV/EHU, Apartado 644, 48080 Bilbao, Spain}
\affiliation{EHU Quantum Center, University of the Basque Country UPV/EHU, Apartado 644, 48080 Bilbao, Spain}
\affiliation{Quantum Mads, Uribitarte Kalea 6, 48001 Bilbao, Spain}
	
\author{Lucas Lamata}
\affiliation{Departamento de F\'isica At\'omica, Molecular y Nuclear, Universidad de Sevilla, 41080 Sevilla, Spain}
\affiliation{Instituto Carlos I de F\'isica Te\'orica y Computacional, 18071 Granada, Spain}

\author{Jos\'e D. Mart\'in-Guerrero}
\affiliation{IDAL, Electronic Engineering Department, University of Valencia, Avgda. Universitat s/n, 46100 Burjassot, Valencia, Spain ES}
\affiliation{ValgrAI:Valencian Graduated School and Research Network of Artificial Intelligence, Spain ES}
\author{Enrique Lizaso}
\affiliation{Multiverse Computing, Pio Baroja 37, 20008 San Sebasti\'an, Spain}
\author{Samuel Mugel}
\affiliation{Multiverse Computing, Pio Baroja 37, 20008 San Sebasti\'an, Spain}
\author{Xi Chen}
\affiliation{Department of Physical Chemistry, University of the Basque Country UPV/EHU, Apartado 644, 48080 Bilbao, Spain}
\affiliation{EHU Quantum Center, University of the Basque Country UPV/EHU, Apartado 644, 48080 Bilbao, Spain}
\author{Rom\'an Or\'us}
\affiliation{Multiverse Computing, Pio Baroja 37, 20008 San Sebasti\'an, Spain}
\affiliation{Donostia International Physics Center, Paseo Manuel de Lardizabal 4, 20018 San Sebasti\'an, Spain }
\affiliation{IKERBASQUE, Basque Foundation for Science, Plaza Euskadi 5, 48009, Bilbao, Spain}
\author{Enrique Solano}
\affiliation{International Center of Quantum Artificial Intelligence for Science and Technology (QuArtist) \\ and Department of Physics, Shanghai University, 200444 Shanghai, China}
\affiliation{IKERBASQUE, Basque Foundation for Science, Plaza Euskadi 5, 48009, Bilbao, Spain}
\affiliation{Kipu Quantum, Greifswalderstrasse 226, 10405 Berlin, Germany}
\author{Mikel Sanz}
\email{mikel.sanz@ehu.eus}
\affiliation{Department of Physical Chemistry, University of the Basque Country UPV/EHU, Apartado 644, 48080 Bilbao, Spain}
\affiliation{EHU Quantum Center, University of the Basque Country UPV/EHU, Apartado 644, 48080 Bilbao, Spain}
\affiliation{IKERBASQUE, Basque Foundation for Science, Plaza Euskadi 5, 48009, Bilbao, Spain}
\affiliation{Basque Center for Applied Mathematics (BCAM),  Alameda de Mazarredo, 14, 48009 Bilbao, Spain}

\begin{abstract}
Prediction of financial crashes in a complex financial network is known to be an NP-hard problem, which means that no known algorithm can guarantee to find optimal solutions efficiently. We experimentally explore a novel approach to this problem by using a D-Wave quantum annealer, benchmarking its performance for attaining a financial equilibrium. To be specific, the equilibrium condition of a nonlinear financial model is embedded into a higher-order unconstrained binary optimization (HUBO) problem, which is then transformed to a spin-$1/2$ Hamiltonian with at most two-qubit interactions. The problem is thus equivalent to finding the ground state of an interacting spin Hamiltonian, which can be approximated with a quantum annealer. The size of the simulation is mainly constrained by the necessity of a large quantity of physical qubits representing a logical qubit with the correct connectivity. Our experiment paves the way to codify this quantitative macroeconomics problem in quantum annealers.
\end{abstract}
\maketitle
\section{Introduction}

Economics is a complex science in which the agents’ psychology plays an essential role that is often hardly grasped by mathematical models. However, economists do not relinquish to try to predict market behavior employing sophisticated models, leading to the field of quantitative finance. Following this idea, quantitative finance and economics emerged. They were applied to understand the evolution of financial markets and economies, as well as to forecast their possible future. A realistic question in risk management is: would there be a drastic drop in the market values if the prices of assets suffer some small perturbation? The cross-holdings and nonlinear character of financial network dynamics will cause chain reactions, implying that sudden drops of a market value might affect other nodes in the network resulting in a financial crisis. Presently, the prediction of crashes is mainly performed by studying previous cases in history and comparing with the current configuration~\cite{prediction1,prediction2,prediction3,prediction4,prediction5,prediction6}. While this empirical approach has been succesful~\cite{IEEE-classicalreview}, the economic environment is constantly evolving. Hence, we cannot limit ourselves to predicting economic disasters which are qualitatively similar to past events. Therefore, ab initio simulations of financial networks will become essential for avoiding financial crises. This problem was recently shown to be NP-Hard~\cite{np}.  Therefore, given the current standpoint on complexity theory, this problem is not expected to be efficiently solvable in a classical computer. Indeed, given the global knowledge of a financial network, the time to compute the consequences of a perturbation would by far exceed the age of the universe.

An alternative approach to this problem is presented in Refs.~\cite{crash,usecases}, where they depict how to tackle this type of problems by using quantum annealers. In particular, a mathematically identical problem is simulated and the corresponding result measured~\cite{annealing1,annealing2,annealing3,annealing4}. Specifically, it was shown that obtaining the equilibrium configuration of a financial network is equivalent to solving a higher-order unconstrained binary optimization (HUBO) problem, which should be feasible for a quantum annealer allowing for multi-qubit interactions. Unfortunately, this hardware has not been realized yet, as state-of-the-art quantum annealers are restricted to two-qubit interactions~\cite{evidence}. A possible work-around, which comes at the price of introducing ancillary qubits, is to find an effective Hamiltonian with the same low-energy subspace and two-qubit interactions at most. This leaves us with the problem of solving a quadratic unconstrained binary optimization (QUBO) problem whose optimum encodes the equilibrium configuration of a financial network. This problem can be addressed employing a quantum annealer. The D-Wave 2000Q quantum annealer, equipped with a Chimera architecture, requires a large quantity of physical qubits to obtain the desired connectivity and limits the number of institutions and assets considered. An analysis of the changes experienced by the financial network to reach its equilibrium configuration will tell whether a crash has occurred.

In this paper, we experimentally implement the study presented in Refs.~\cite{crash,usecases}. Specifically, we compute the equilibrium configuration of a financial network before and after perturbation with a D-Wave 2000Q quantum annealer, and compare the result to alternative methods. Although the D-Wave machine has been successfully used to solve problems in engineering~\cite{volkswagen}, cryptography~\cite{FengHu}, biology~\cite{bio}, and quantitative finance~\cite{quantfin1,quantfin2} among others, it is the first time that quantum annealing is applied to solve a problem of macroeconomics. This should attract more attention from the finance and economic disciplines towards quantum computing~\cite{nielsenchuang,QuantumFinance,Dwave1,Dwave2,Dwave3,Dwave4,Dwave5}. 

The contents are organized as follows: in Sec.~\ref{sec2}, we introduce the model of financial network that will be considered. Sec.~\ref{sec3} reviews the quantum annealing algorithm to find financial equilibrium. Sec.~\ref{sec4} experimentally proves the validity of the scheme by finding the financial equilibrium of a random network of the largest implementable size with a D-Wave 2000Q quantum annealer; for this network, we also show experimentally how the scheme allows to compute the financial equilibrium. Sec.~\ref{sec5} analyzes the achieved results and discusses further possible improvements. The conclusions drawn from the work are shown in Sec.~\ref{sec6}.

\section{Formulation of the model}
\label{sec2}

\begin{figure}[b]
\centering
\includegraphics[scale=0.2]{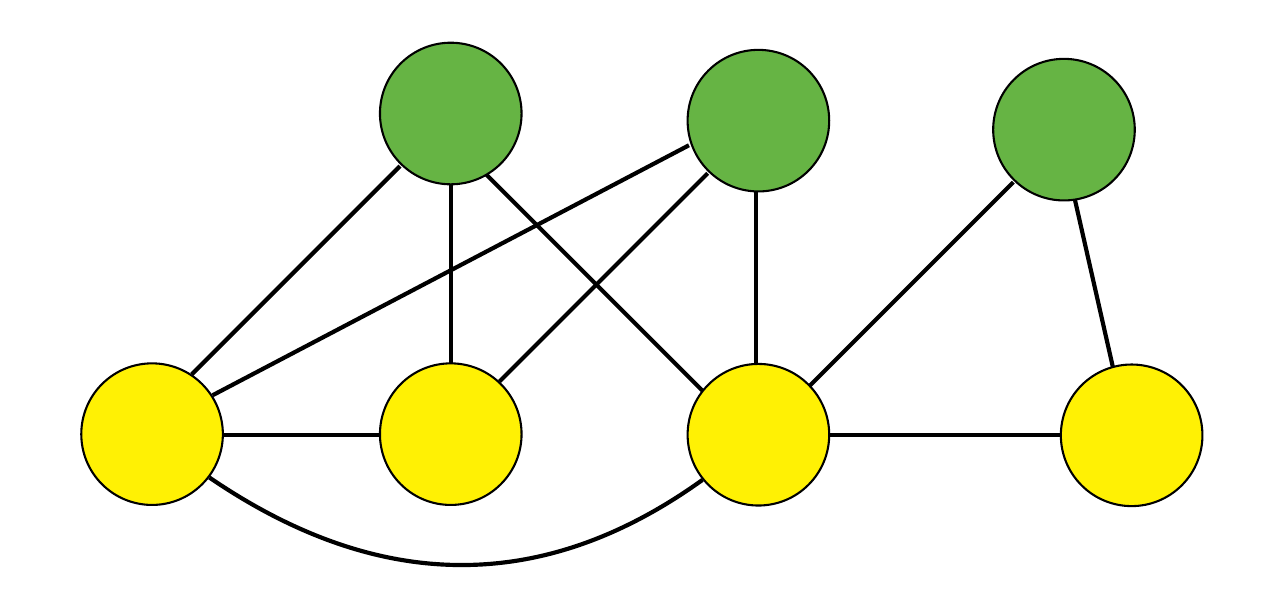}
\caption{\label{network} Example of a financial network: the yellow nodes and green nodes denote institutions and assets, respectively. Links denote ownerships and cross-holdings, described by the ownership matrices ${\bf D}$ and ${\bf C}$, respectively. Diagonal matrix $\tilde{{\bf C}}$ represents the self-ownership of institutions, which would be plotted as self-loops in the graph representation. The equity value $V_i$ of institution $i$ is defined by summing up its ownership of all assets and cross-holdings. }
\end{figure}

A nonlinear network model for financial markets is proposed in  Ref.~\cite{crash}. It is made up of $n$ institutions and $m$ assets, and aims at representing the market values of institutions by mapping it onto a graph, as shown in Fig.~\ref{network}.  We codify the prices of the $m$ assets by an $m-$dimensional vector $\vec{p} \in \mathbb{R}^m$, where the element $p_k$ represents the price of asset $k$. Moreover, an  $n\times m$ ownership matrix ${\bf D}$ can be defined such that the element $D_{ik}\geq0$ corresponds to the percentage of asset $k$ owned by institution $i$. There is also an $n\times n$ ownership matrix $\mathcal{C}$ that describes the cross-holdings and self-ownerships between institutions. The coefficients $\mathcal{C}_{ij}$ denote the percentage of institution $j$ owned by institution $i$. By considering all self-ownerships (i.e., the diagonal elements) from $\mathcal{C}$ one forms a new diagonal matrix $\tilde{{\bf C}}$ which represents the self-ownership only, such that matrix ${\bf C}=\mathcal{C}-\tilde{{\bf C}}$ codifies all cross-holdings. The equity value $V_i$ of institution $i$ is defined by summing up its ownership of all assets and cross-holdings, $V_i=\Sigma_kD_{ik}p_k+\Sigma_jC_{ij}V_j$. One thus obtains a matrix equation $\vec{V}={\bf D}\vec{p}+{\bf C}\vec{V}$, where equity value vector $\vec{V }\in \mathbb{R}^n$ is an $n-$dimensional vector. Accordingly, the market value is the equity value rescaled with its self-ownership, resulting in the $n-$dimensional market value vector $\vec{v}=\tilde{{\bf C}}\vec{V}$. The solution to the linear matrix equation thus reads
\begin{equation}
\vec{v}=\tilde{{\bf C}}({\bf I}-{\bf C})^{-1}{\bf D}\vec{p}.
\label{eq1}
\end{equation}

We introduce the nonlinear effect of \emph{panic} in the model via a Heaviside-theta function $\Theta$; if the market value $v_i$ drops below the critical value $v^i_c$, \emph{failure} of institution $i$ occurs and its equity value drops by $\beta_i(\vec{p})$ which is governed by the price vector of assets. Once we define the failure vector $\vec{b}(\vec{v},\vec{p})=\vec{\beta}(\vec{p})\circ(1-\Theta(\vec{v}-\vec{v_c}))$, where $\circ$ denotes the Hadamard product, the market value vector with nonlinearity can be written as
\begin{equation}
\label{eq-non}
\vec{v}=\tilde{{\bf C}}({\bf I}-{\bf C})^{-1}({\bf D}\vec{p}-\vec{b}(\vec{v},\vec{p})).
\end{equation}
Mathematically, it is the nonlinearity of $\vec{b}(\vec{v},\vec{p})$ which makes financial networks so hard to be predicted. 
This drop may cause an institution's value to \emph{crash}, a behavior which can infect other nodes in the network. Under our definition, a financial crash happens when, after a perturbation in the assets price, the market value of an institution considering the nonlinear term is lower than those pre-perturbation prices calculated with the linear model.

\section{Quantum annealing algorithm}
\label{sec3}
As proposed in Ref.~\cite{crash} finding financial equilibrium can be represented as the minimization of an objective function, which is equivalent to finding the ground state of a classical spin Hamiltonian.

By squaring  Eq.~(\ref{eq-non}), we obtain an objective function that meets its minimum value when the market value state is set to be the equilibrium state
\begin{equation}
\label{eq:objective}
\text{Obj}(\vec{v})=(\vec{v}-\tilde{{\bf C}}({\bf I}-{\bf C})^{-1}({\bf D}\vec{p}-\vec{b}(\vec{v},\vec{p})))^2.
\end{equation}
Thus, our task is now to find the $\vec{v}$ that minimizes $Obj(\vec{v})$ for a given financial network, which is an NP-Hard problem \cite{NPHard}.

Next, we need to deal with the nonlinear terms (modeling failure) of the objective function. The reason is that once the objective function is transformed to a spin-$1/2$ Hamiltonian, it should ideally be made of polynomial terms only, due to the limitations of quantum annealers. Thus, one expands the failure terms with Heaviside-theta functions in terms of polynomials. This expansion is not unique, and here we choose the Legendre expansion~\cite{crash},
\begin{equation}
\label{eq:poly_exp}
\Theta(x)=\frac{1}{2}+\sum_{l=1}^{\infty}(P_{l-1}(0)+P_{l+1}(0))P_l(x),
\end{equation}
in the domain $[-1,1]$, with $P_l(x)$ to be the $l$-th Legendre polynomial. By setting $x=(v_i-v^c_i)/v^i_{max}$, Eq.~\eqref{eq:poly_exp} enables us to expand $\Theta(v_i-v^c_i)$ in the range of $v_i\in[0,v^i_{max}]$. Using this expansion as an example, we take the approximation 

\begin{equation}
b_i(v_i,\vec{p})\approx \beta_i(\vec{p}) \left(\frac{1 }{2} -\sum_{l=0}^{r} \Gamma_l 2^l \sum_{k=0}^l \binom{l}{k}\binom{\frac{l+k-1}{2}}{l} \bar{v}^k_i\right)
\end{equation}

where $\Gamma_l=\frac{\sqrt{\pi}}{2\Gamma\left(\frac{2-l}{2}\right)\Gamma\left(\frac{3+l}{2}\right)}$ and $\bar{v}_i=\frac{v_i-v_i^c}{v^i_{\text{max}}}$.The polynomial expansion removes the discontinuity while maintaining the strong nonlinearity of the network.

We now encode the continuous variables $v_i$ with classical bits. This will allow rewriting the resulting objective function into digital form. The expansion is straightforward, and reads $v_i=\sum_{\alpha=-\infty}^{\infty}x_{i,\alpha}2^\alpha$. However, due to the limited resources in real-world devices, one must truncate this expansion, i.e., $v_i\approx\sum_{\alpha=-q}^{q}x_{i,\alpha}2^\alpha$, where $x_{i,\alpha}$ are classical bits with binary values $0$ or $1$. In this way, the market value of institution $i$ is encoded with $2q+1$ classical bits. The maximal market value $v_i^{max}$ is given by $\sum_{\alpha=-q}^q2^\alpha$.

Considering  $(v_i-v_i^c)^k= \sum_{h=0}^k (-1)^h \binom{k}{h} v_i^{k-h} (v_i^c)^h$
and $v_i^{n}= \sum_{m_0+...+m_p=n} \binom{n}{m_0, ..., m_p}  \prod_{0\leq\alpha\leq p}^{m_{\alpha}\neq 0} 2^{\alpha m_{\alpha}} x_{i,\alpha}$,
the resulting objective function is a polynomial of binary variables $x_{i,\alpha}$ of degree $2r$.

\begin{equation}
\hat{H}=\sum_i\left( \sum_{\alpha=-q}^{q}x_{i,\alpha}2^\alpha - \gamma_i + \sum_j \bar{C}_{ij} b_j (x_{j,\alpha},\vec{p})\right)^2
\end{equation}
with $\gamma_i=\sum_j\bar{C}_{ij} \sum_k D_{jk}p_k  $ and $\bar{C}_{ij}=\tilde{C_{ii}} (\textbf{I}-C)^{-1}_{ij}$. To express it as a spin-1/2 Hamiltonian, we replace the binary variables $x_{i,\alpha}$ by qubit operators $\hat{x}_{i,\alpha}$ with eigenvalues  $0$ and $1$, i.e., $\hat{x}_{i,\alpha}|0\rangle=0$, $\hat{x}_{i,\alpha}|1\rangle=|1\rangle$. The Pauli-$z$ operator satisfies $\hat{x}_{i,\alpha}=(1+\hat{\sigma}^z_{i,\alpha})/2$, and therefore the Hamiltonian encodes the objective function but written with Pauli matrices, including all types of multi-spin interactions, up to $2r$-body terms.
\begin{figure}[b]
\centering
\includegraphics[scale=0.9]{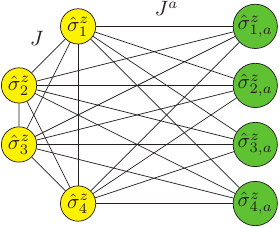}
\caption{\label{k-2} Recast of quantum Hamiltonian with $k$-qubit interactions into a modified, effective Hamiltonian with the same low-energy spectrum with two-qubit interactions at most. We illustrate the particular case of a $k=4$-qubit interactions, which requieres the introduction of  $4$ ancilla qubits to obtain the effective Hamiltonian.}
\end{figure}
The Hamiltonian obtained is appropriate for a quantum annealer that allows many-qubit interactions. However, state-of-the-art quantum annealers only accept inputs with at most two-qubit interactions. Finding the ground state of a spin-1/2 Hamiltonian is equivalent to solving a Quadratic Unconstrained Binary Optimization (QUBO) problem, which is the input of the quantum annealer. Thus, we should recast our quantum Hamiltonian into a modified, effective Hamiltonian with at most two-qubit interactions. Some protocols achieving exactly this are proposed in Refs.~\cite{many-8,many-7,many-6,many-5,many-4,many-3,many-2}, in particular we base our protocol in Ref.~\cite{many-2}, where $k$ ancilla qubits are introduced to implement an effective $k$-qubit interaction. Suppose that there is a $k$-qubit interaction term $\hat{H}_k=J_k\Pi_{i=1}^k\sigma^z_{i}$ with the same low-energy spectrum of another Hamiltonian term $\tilde{H}_k$ with at most two-qubit interactions. We can express $\tilde{H}_k$ with $k$ logical qubits and $k$ extra ancilla qubits as
\begin{eqnarray}
&&\tilde{H}_k=J\sum_{i=2}^k\sum_{j=1}^{i-1}\hat{\sigma}^z_i\hat{\sigma}^z_j+h\sum_{i=1}^k\hat{\sigma}^z_i\nonumber\\&&+J^a\sum_{i=1}^k\sum_{j=1}^k\hat{\sigma}^z_i\hat{\sigma}^z_{j,a}+\sum_{i=1}^kh_i^a\hat{\sigma}^z_{i,a} ,
\label{twoqhamil}
\end{eqnarray}
as represented in Fig.~\ref{k-2}. This two-qubit Hamiltonian has the same low-energy spectrum than $\hat{H}_k$ when $J$, $J^a$, $h$ and $h_i^a$ are set to some appropriate values.  As Ref.~\cite{many-2} suggested, this can be achieved once $q_i=(-1)^{k-i+1}J_k+q_0$, $h=-J^a+q_0$, $h_i^a=-J^a(2i-k)+q_i$ and $J=J^a$, with any $q_0$ that satisfies $|J_k|\ll q_0<J^a$ and $|J_k|\ll J^a-q_0<J^a$. These conditions can be relaxed to $|J_k|<q_0<J^a$ and $|J_k|<J^a-q_0<J^a$ if one aims at having the same ground state only, instead of the whole low-energy sector. We depict the low-energy spectrum of this two-qubit Hamiltonian for $k=3$ logical qubits in Table \ref{tb}.

\begin{table}[h!]
\centering
\begin{tabular}{cccccc||c}
 $\hat{\sigma}_1$& $\hat{\sigma}_2$& $\hat{\sigma}_3$& $\hat{\sigma}^a_1$& $\hat{\sigma}^a_2$& $\hat{\sigma}^a_3$&Energy (u) \\
\hline
\hline
1 & 1  & -1 &-1 & -1  & 1 & -121 \\ \hline
1 & -1  & 1 &-1 & -1  & 1 & -121 \\ \hline
-1 & 1  & 1 &-1 & -1  & 1 & -121 \\ \hline
-1 & -1  & -1 &1 & 1  & 1 & -121 \\ \hline 
1 & 1  & 1 &-1 & -1  & -1 & -119 \\ \hline
1 & -1  & -1 &-1 & 1  & 1 & -119  \\ \hline
-1 & 1  & -1 &-1 & 1  & 1 & -119  \\ \hline
-1 & -1  & 1 &-1 & 1  & 1 & -119  \\\end{tabular}
\caption{Low energy spectrum (first 8 eigenstates) of the two-qubit Hamiltonian, Eq.(\ref{twoqhamil}) with $k=3$, result of mapping the term  $\hat{\sigma}_1\hat{\sigma}_2\hat{\sigma}_3$ according to Ref.~\cite{many-2}. Value of the parameters $J_3$=1 \ \text{u}, $J=J_a=20 \ \text{u}$, $q_0=10\ \text{u}$, $h=-10 \ \text{u}$, $q_1=q_3=9 \ \text{u}$, $q2=11 \ \text{u}$,  $h_1=29 \ \text{u}$, $h_2=-9 \ \text{u}$, $h_3=-51 \ \text{u}$. }
\label{tb}
\end{table}

\newpage
\section{Implementation in a D-Wave 2000Q quantum annealer}
\label{sec4}

\begin{figure}[b]
\centering
\includegraphics[scale=0.8]{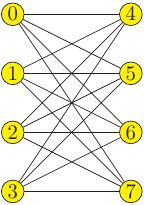}
\caption{\label{chimera} Chimera graph topology implemented by the D-Wave 2000Q quantum annealer. The 2048 qubits are partitioned into subgraphs of 8 qubits. The connection between subgraphs is sparse, in each of these subgraphs there are two sets of four qubits; each qubit connects to all qubits in the other set but to none in its own, forming a ${\bf K_{4,4}}$ bipartite graph.}
\end{figure}

Once shown that it is possible to recast the problem of finding financial equilibrium into a language amenable to QUBO solvers, and in particular, quantum annealers, this section deals with its implementation using a state-of-the-art quantum annealer, namely, the D-Wave 2000Q. This quantum annealer consists of up to 2048 qubits connected according to the Chimera graph topology (see Fig.~\ref{chimera}). It is designed to solve embedded Ising problems or QUBO problems.

Two simulations were produced:
\begin{enumerate}
 \item  A financial network without failure term, which is simple to solve on a classical computer in order to benchmark the performance of the quantum processor.
 \item A financial network with the inherently nonlinear risk of failure. We will perturb the asset price vector in this network to compute the new equilibrium configuration using the quantum annealing algorithm.
\end{enumerate}

\begin{figure*}[t]
\centering
\includegraphics[width=1.6\columnwidth]{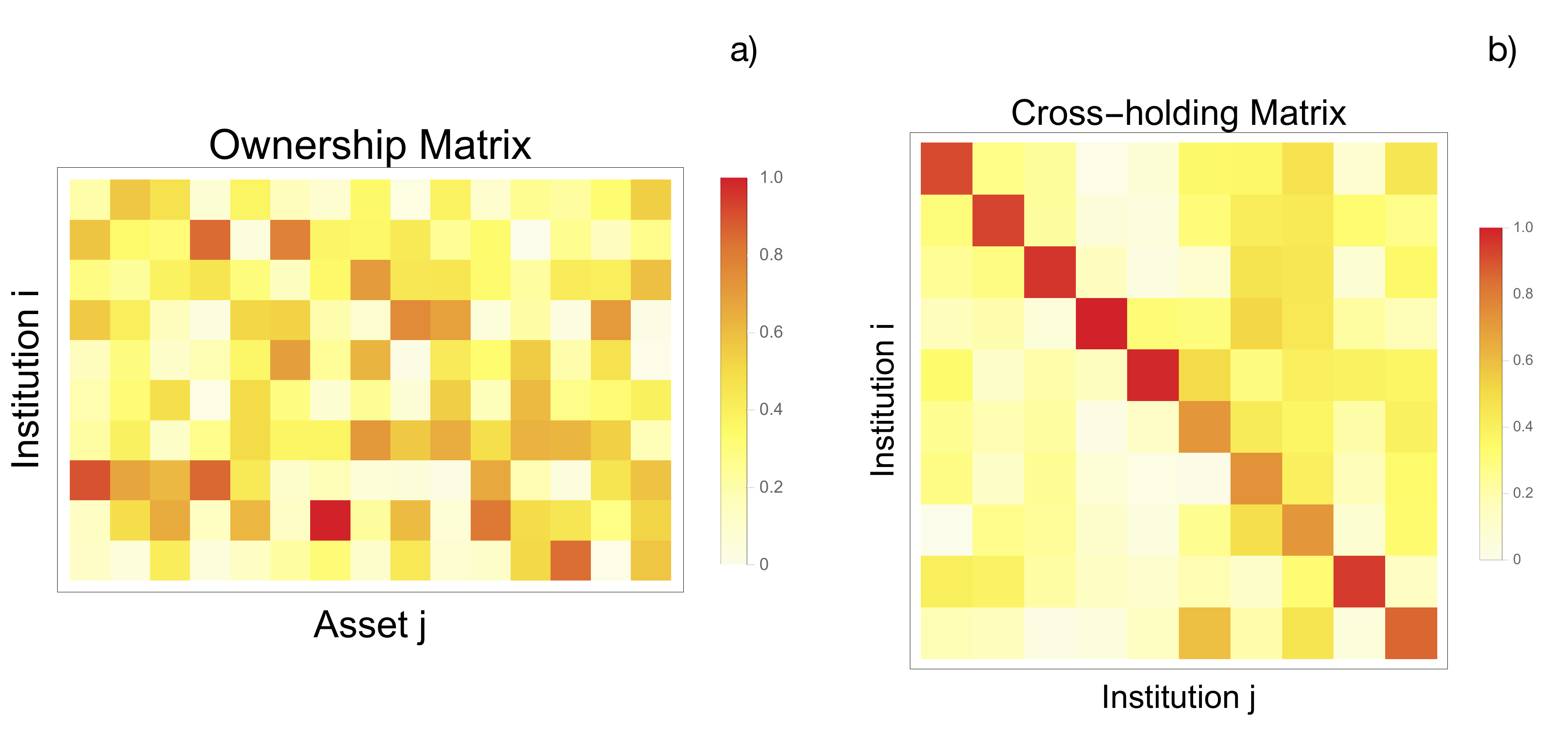}
\caption{a) Ownership matrix ${\bf D}$ for the linear model. The element $D_{ik}\geq0$ corresponds to the percentage of asset $k$ owned by institution $i$. We randomize the ownership matrix ${\bf D}$ with a Dirichlet distribution that satisfies $\sum_{i=1}^n D_{ij}=1$. b)  Cross-holding matrix $\mathcal{C}$ for the linear model that describes the cross-holdings and self-ownerships between institutions.  Cross-holding matrix is generated in a similar way to ownership matrix but with a constraint that all diagonal elements should be larger than $0.5$, ensuring that all institutions can make decisions according to their own wills. All of these data, as well of the asset prices, have been synthetically produced but following all constraint conditions proposed in the theoretical model \cite{crash}.  }
\label{C}
\end{figure*}

We initially generate a financial network with $10$ institutions and $15$ assets. To demonstrate the algorithm, we randomize the ownership matrix ${\bf D}$ with a Dirichlet distribution that satisfies $\sum_{i=1}^n D_{ij}=1$, where $D_{ij}$ are random variables. The cross-holding matrix $\mathcal{C}$ is generated in a similar way but with a constraint that all diagonal elements should be larger than $0.5$, ensuring that all institutions can make decisions according to their own wills. Thus, we randomize $\tilde{C}_{ii}$ between $0.5$ and $1$ and randomize $\sum_{i=1}^nC_{ij}=1-\tilde{C}_{jj}$ with a rescaled Dirichlet distribution. The price vector $\vec{p}$ is also random, with $p_i\in[10,40]$. The network configuration is shown in Figs.~\ref{C} a) and~\ref{C} b).

We can calculate the equilibrium state $\vec{v_q}$ and the equity value vector $\vec{V}$ on a classical computer using
\begin{eqnarray}
\vec{v_q}=\tilde{{\bf C}}({\bf I}-{\bf C})^{-1}{\bf D}\vec{p},\\
\vec{V}=({\bf I}-{\bf C})^{-1}{\bf D}\vec{p}.
\end{eqnarray}
which are linear equations that in fact can be implemented in a quantum annealer by using only 2-local terms, result of squaring the expression in a similar way to Eq.~\eqref{eq:objective}.

The objective function shown in Eq.~\eqref{eq:objective} was implemented, for benchmarking reasons, both in a quantum annealer and a classical simulator. Variables $v_i$ were encoded, $v_i=\sum_{\alpha=0}^62^\alpha x_{i,\alpha}$, on seven qubits. As such, this constrains the $v_i$ to be integers smaller than 127. A quantum implementation of this algorithm does not require ancilla qubits, as there are no many-qubit interactions.

\begin{figure}[b]
\centering
\includegraphics[scale=0.7]{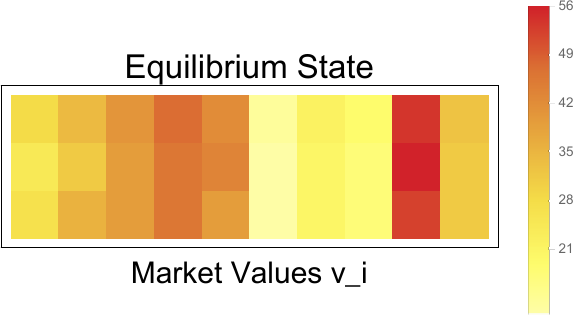}
\caption{\label{vlinear} Linear-model result. The first row shows the result if the matrix equation is solved exactly, the second row if \textit{qbsolv} with tabu classical solver is used, and the third row if \textit{qbsolv} with D-Wave 2000Q solver is employed. We observe that comparing the individual equilibrium values, a quantum annealer provides a compatible solution to the exact solution.}
\end{figure}

The QUBO for this linear problem is a $70\times70$ matrix, with 210 couplers which cannot be solved directly due to the topology structure of the quantum annealer. D-Wave provides a software named {\it qbsolv} that allows to combine the classical computer with its quantum annealer by splitting the QUBO matrix into partition matrices that can be embedded in the quantum annealer. As a decomposing solver, it finds a minimum value of a large QUBO problem by splitting it into pieces solved either via a D-Wave system or a classical tabu solver (both approaches were considered here for comparison purposes). Since the D-Wave 2000Q processor is a quantum annealer, $20$ results would be obtained from a {\it qbsolv} process with a default setting; these results should be handled by a correction process, e.g.,  majority voting, to help us identify the most plausible answer. The result of this QUBO problem is shown in Fig.~\ref{vlinear}, where the exact solution via solving a linear matrix equation, {\it qbsolv} solution with classical tabu solver, and {\it qbsolv} solution with D-Wave quantum annealer, are compared. It is straightforward to observe that comparing the individual equilibrium values, a quantum annealer provides a solution that present more accurate individual values of the assets than the prediction from the classical solver. 

While the failure-free model only has linear and quadratic terms in $v_i$, the nonlinear model has powers of $v_i$ up to order $2r$. For large $r$, this can be extremely resource-consuming in terms of ancillary qubits due to the requested connectivity. An estimation of the number of qubits can be made by counting the number of interaction terms; Our Hamiltonian $\hat{H}$ can have up to $\sum_{\alpha=0}^{2r}\ \binom{n(2q+1)}{\alpha}$ terms, where $n(2q+1)$ denotes the logical qubits that are required. In each term, $3$-to-$2r$ new ancilla qubits are needed, depending on the number of logical qubits in this term. Therefore, the number of necessary qubits grows rapidly with the degree of the polynomial expansion $r$. Note that the aforementioned QUBO problem is NP-hard for any $n\geq2$. In practice, this is an upper bound to the required resources, calculated assuming that $\hat{H}$ has all possible terms up to order $O(2r)$.

\begin{figure*}[t]
\centering
\includegraphics[width=1.6\columnwidth]{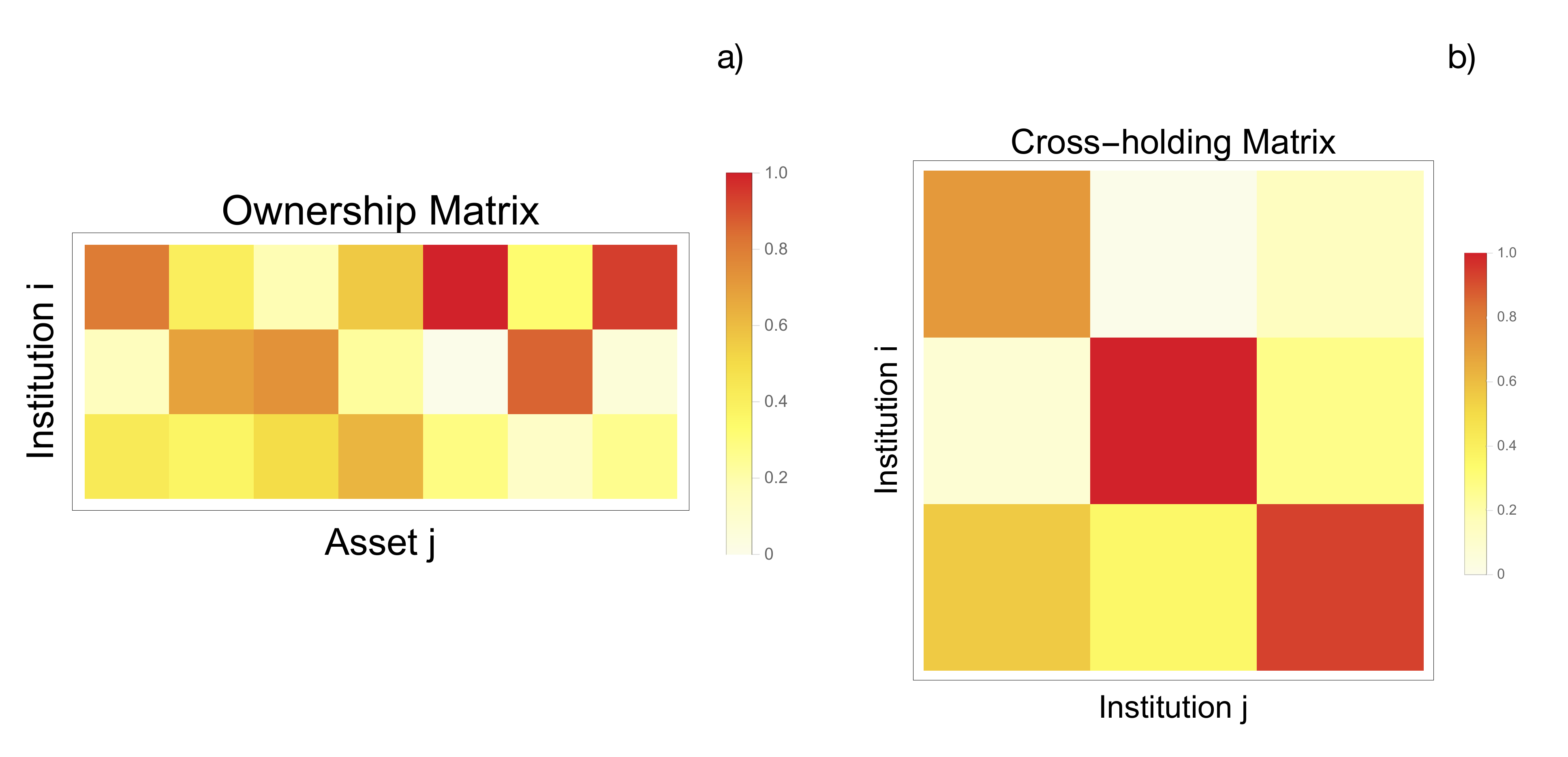}
\caption{a) Ownership matrix ${\bf D}$ for the implemented network with failure terms. The element $D_{ik}\geq0$ corresponds to the percentage of asset $k$ owned by institution $i$. We randomize the ownership matrix ${\bf D}$ with a Dirichlet distribution that satisfies $\sum_{i=1}^n D_{ij}=1$. b) Cross-holding matrix $\mathcal{C}$ for the implemented network with failure terms that describes the cross-holdings and self-ownerships between institutions. Cross-holding matrix is generated in a similar way to ownership matrix but with a constraint that all diagonal elements should be larger than $0.5$, ensuring that all institutions can make decisions according to their own wills. }
\label{newD}
\end{figure*}

Here, we implement an enhanced model with failure terms on the basis of the linear model previously simulated. We perturb the vector of asset prices, leaving the ownership matrix ${\bf D}$ and cross-holding matrix $\mathcal{C}$ invariant, and recompute the equilibrium state. Specifically, we set the price of some random assets to zero (to simulate, e.g: the assets' destruction). In this study, we will use an expansion of $\hat{H}$ to third order, which still characterizes the phenomenon of sudden drop near the critical value. Moreover, this approach provides strong nonlinearity while saving plenty of qubit resources. As a result, 70 logical qubits and 872,690 ancilla qubits are required, which leads to a QUBO matrix of 872,760x872,760 entries, although only a minority of them, 4,446,575 couplers, are non-zero. Storing this sparse matrix results in the requirement of about 6TB RAM memory, since each element has an accuracy of double float in {\it qbsolv}. Due to the limitations of state-of-the-art techniques, the network is reduced to three institutions and each market value $v_i$ is encoded by five qubits, bounding the maximum market value to be $31$. New $3\times7$ ownership matrix ${\bf D}$ and $3\times3$ cross-holding matrix $\mathcal{C}$ are generated while the price vector $\vec{p}$ before perturbation is $\vec{p}=\{8.43,\ 14.47,\ 6.75,\ 8.09,\ 19.11\ ,11.32,\ 7.19\}^T$. The network configuration is shown in Figs.~\ref{newD} a) and~\ref{newD} b). The equilibrium state before perturbation without nonlinearity is given as $\vec{v_q}=\{21.18\,\ 23.33,\ 30.83\}^T$, and the critical value vector is still set to be $80\%$ of the original equilibrium state, while the failure strength $\vec{\beta}$ is considered to be $30\%$ of the original equity value. The corresponding perturbed price vector is given as $\vec{p}=\{8.43,\ 14.47,\ 0,\ 8.09,\ 0\ ,11.32,\ 7.19\}^T$. Before calculating the new equilibrum state with nonlinearity and perturbation, some parameters, like $J^a$ and $q_0$, must be set. For the minor embedding of a submatrix in the D-Wave quantum annealer, this is done by introducing a penalty function between qubits in the Chimera graph requiring $J^m\geq J^a$, which means that the $J^a$ for mapping multi-qubit interactions to two-qubit interactions should be in the proper scale. Meanwhile, as we mentioned in the theory part, we need to sample out the thermal fluctuation by assuming that $|\hat{H}_k|$ is much smaller than $J^a$, or the protocol will break down because those ancilla qubits will not be in the corresponding ground state anymore. Thus, in the implementation we took $J^a=20J_k$ and $q_0=10J_k$, such that this could ensure that either $q_0$ or $J^a-q_0$ would be at least 10 times larger than $J_k$.

\begin{figure}[b]
\centering
\includegraphics[width=0.85\columnwidth]{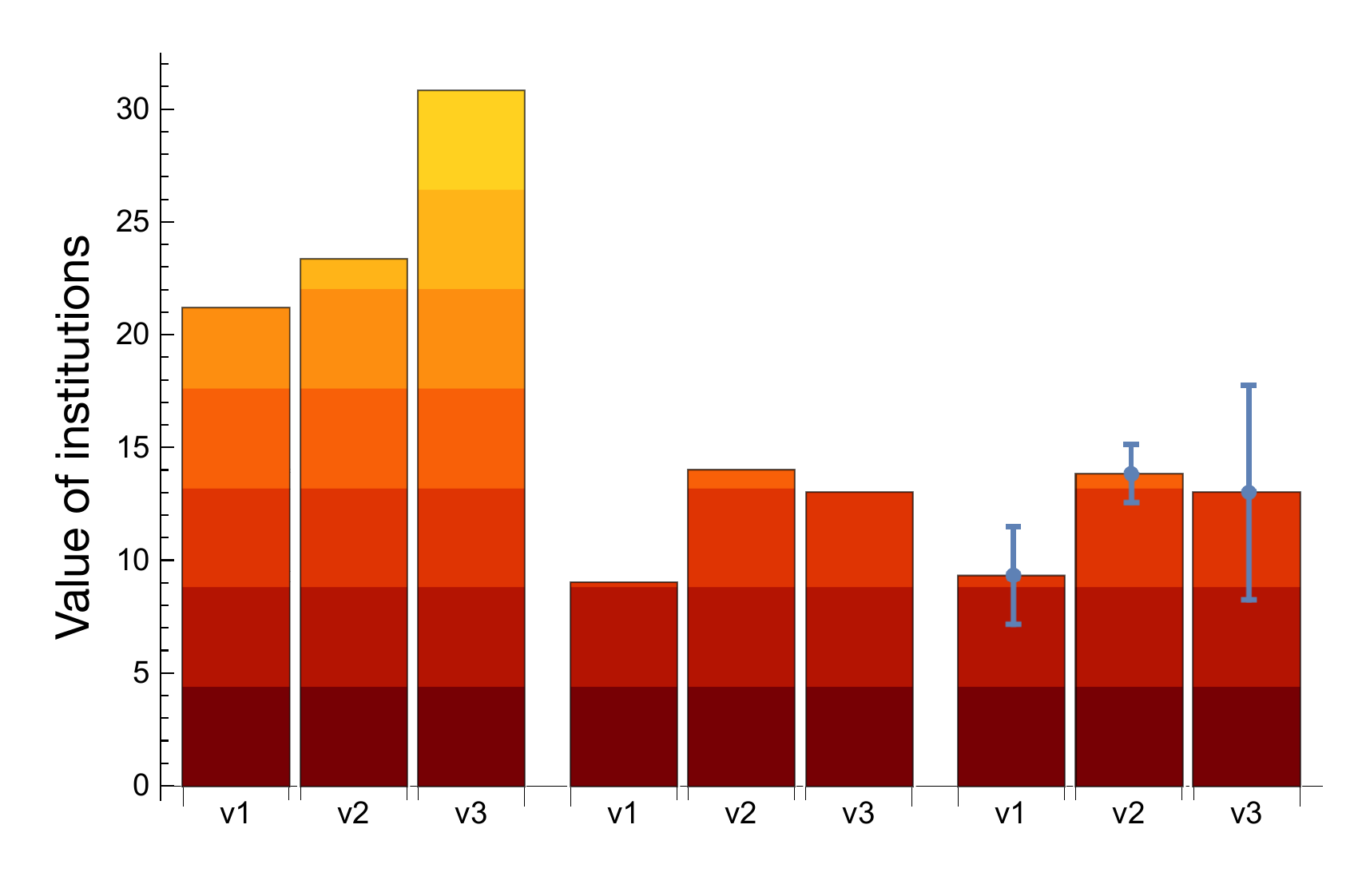}
\caption{Market values $v_1$, $v_2$ and $v_3$ of institutions 1, 2, and 3, respectively for different scenarios. The first group (left) is the equilibrium state without taking non-linearity terms (perturbations) into consideration, where the asset price is calculated by inverting the matrix of Eq. (\ref{eq1}). The second group (center) is the equilibrium state after taking nonlinearity as  the `failure term', which is activated by a critical value vector $80\%$ of the original equilibrium state
calculated with a straightforward method by trying  $32^3$ times by brute force, corresponding to all the possible combinations. The third group (right) shows the outcome from {\it qbsolv} software in D-Wave 2000Q. The error bar characterizes a 95\% confidence interval. The agreement between integer and annealer solutions confirms the feasibility and accuracy of the method. Additionally, by comparing both with the pre-perturbation values, we can conclude that we have detected the financial crash.}
\label{nonlinearv}
\end{figure}

For this problem, the QUBO matrix had the size of $8280\times 8280$, with 15 logical qubits, 8265 ancilla qubits and 38,790 couplers. Remark that the available quantum annealer structure is not optimized for this problem and, also, that the translation is not efficient because of sparse connectivity of the quantum processor.  Finally, we compare our results from the quantum annealer with the integer equilibrium solution calculated with a straightforward method by trying  $32^3$ times in Fig.~\ref{nonlinearv}, which shows a good agreement and the accuracy of the proposed method. Comparing the results after the perturbation with the pre-perturbation values, we can conclude that we have detected the financial crash.

\section{Results and discussion}
\label{sec5}

D-Wave is a quantum annealer designed to deal with Ising Model and QUBO problems. However, the problem faced in this paper, namely, financial crisis prediction with nonlinearity associated to panic, is not QUBO but HUBO instead, thus requiring multi-qubit interactions. In order to approximate this HUBO problem with two-qubit interactions, at the current stage of hardware and software we were limited to simulate a small financial network, made up of three institutions and cross-holdings.

An effective two-qubit quantum Hamiltonian could still not be read directly in D-Wave system which requires QUBO type input or Ising type input. Although some open-source software like \textit{pyqubo} can generate it, the input size must be very small in order to avoid a stack overflow associated with recursion errors. A possible solution is to produce a Mathematica script that reads each term, write it as a string of coefficients and qubits in an input file for the D-Wave system. Once we generate the input for this problem, this is still too large to be embedded in the D-Wave 2000Q quantum annealer because of the graph structure. Thus, {\it qbsolv} is an inevitable option for us, which works by separating the large matrix to submatrices and solve them by a classical tabu solver or D-Wave solver. This kind of hybrid computation provides the possibility to solve the complicated problem but brings some new constraints, namely: (\romannumeral1) {\it Local hardware.} Once the QUBO matrix is provided, {\it qbsolv} allocates dynamic memory before separating it to submatrices with elements of double precision floats, by requiring a size of $8n\times n $ bytes of memory. However, the bottleneck is not the memory size but the performance of CPU since a large QUBO matrix will consume exhaustive CPU time if one needs high accuracy of the optimized result; (\romannumeral2) {\it Algorithm.} Instead of a real quantum annealing process for the whole matrix, {\it qbsolv} provides a tabu algorithm or D-Wave 2000Q quantum annealer for submatrices. The partition strategy for generating submatrices may get stuck in a local minimum instead of the global minimum that quantum annealing guarantees with high probability under ideal conditions, i.e. in absence of decoherence and in the adiabatic limit. Considering that the logical qubits only encode less than $1\%$ in the QUBO matrix, the risk of getting stuck is still high, even if we sample over the thermal distribution or give a huge repeat limitation in the main loop to improve its accuracy. We would have to customize a random seed for the separation, and check the final result manually, to see whether the result is near from the equilibrium. Another option is that one may send the QUBO matrix to the solver many times and average the result to obtain the best solution. (\romannumeral3) {\it Quantum annealer.} The submatrices will be sent to D-Wave 2000Q quantum annealing device for optimization after they are generated by Glover's algorithm~\cite{qbsolv}. In the quantum annealing process, magnetic fields are applied to the processors and the strength should be accurate because $J^k,J^a$ in the QUBO matrix and $J^m$ for the embedding belong to different magnitudes. Any imprecision in the system preparation will cause significant deviation from the correct result.

In this implementation, the accuracy is not especially high, since we are not optimizing the objective function rigourously because the market values are integers $v_i\in[0,v_{max}]$ constrained for the qubits we take to encode them. The computation time is also long, considering that there is a straightforward but equivalent classical algorithm by testing the value of the objective function $32^3$ times by brute force, corresponding to all the possible combinations. Although mapping it to a QUBO problem and optimizing it with a general quantum annealer is not efficient enough for current technology, we believe it is a valuable example of how one can solve an NP-hard problem via quantum computation. With quantum annealers designed for solving HUBO problems that allow the implementation of multi-qubit interactions, we would avoid the overhead of resources and we may obtain a speed up factor in forecasting the behavior of complex financial networks over the use of general purpose annealers. We expect this kind of quantum solver may be available in the near future. Meanwhile, D-Wave has recently announce its next generation of quantum annealers called the Advantage system~\cite{pegasus}. It would consist of more than 5000 qubits connected with each other according to the Pegasus topology. In this manner, one could improve the number of qubits and the connectivity by a factor of $2.5$.

\begin{figure}[h]
\centering
\includegraphics[scale=0.8]{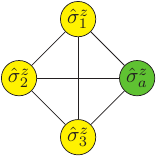}
\caption{\label{3-1} An efficient encoding of three qubits, making use of only one ancilla qubit. The multi-to-two interaction Hamiltonian mapping is a general method, but for three-to-two, a more efficient mapping can be constructed via a subgraph with full connectivity of three logical qubits and one ancilla. }
\end{figure}

Considering that a specialized quantum annealer for HUBO problems would not be available to the public anytime soon, we now analyze the possible ways to enhance the performance of D-Wave 2000Q quantum annealer on this problem. After compromising on the maximum two-qubit interactions in hardware, the subsequent strategy will be reducing the number of ancilla qubits. With fewer ancilla qubits, the size and accuracy of a solvable network can be improved. As proposed in Ref.~\cite{many-2}, the multi-to-two mapping is a general method, but for three-to-two, for example, a more efficient mapping can be constructed with only one ancilla qubit. Suppose there is a sub-Hamiltonian of three-qubit interactions
\begin{equation}
\hat{H}_3=J_3\hat{\sigma}^z_1\hat{\sigma}^z_2\hat{\sigma}^z_3.
\end{equation}
A subgraph with full connectivity of three logical qubits and one ancilla qubit is shown in Fig.~\ref{3-1}, where the equivalent Hamiltonian is given as
\begin{equation}
\tilde{H}_3=J\sum_{i=2}^3\sum_{j=1}^{i-1}\hat{\sigma}^z_i\hat{\sigma}^z_j+h\sum_{i=1}^3\hat{\sigma}^z_i+J^a\sum_{i=1}^3\hat{\sigma}^z_i\hat{\sigma}^z_a+h^a\hat{\sigma}^z_a.
\end{equation}
At variance with the previous protocol, $J^a=2J>h$ and $h^a=2h=2J_3$. Also, for sampling out the thermal fluctuation, we take $J^a\geq J_3$, to prevent the protocol to fail for the same reason. The ancilla qubits can be reduced to about 7000 with this method. Meanwhile, the partition method in {\it qbsolv} may cause the system to get stuck in local minima which requires a better algorithm in the main loop.

\section{Conclusion}
\label{sec6} We have implemented in a D-Wave quantum annealer  the algorithm proposed in Ref.~\cite{crash}, to solve the equilibrium state of a complex financial network that predicts financial crashes. Although the size of the studied financial network is limited, this proof of principle is in agreement with the result of an exhaustive search. This result may be improved with the design of a customized ``financial quantum annealer'': a quantum processor with suitable connectivity for efficient embedding of this kind of problems. Such coherent quantum annealers can be built with current technology~\cite{coherentDWave,LechnerSciAdv,Lechner}, providing convenient multi-qubit couplings.

\begin{acknowledgements}
The authors acknowledge financial support from the project grant PID2021-125823NA-I00 funded by MCIN/AEI/10.13039/501100011033 and by
“ERDF A way of making Europe” and “ERDF Invest in your Future”, Basque Government through Grant No. IT1470-22, the QUANTEK project from ELKARTEK program (KK-2021/00070), as well as from QMiCS (820505) and OpenSuperQ (820363) of the EU Flagship on Quantum Technologies, Spanish Ram\'on y Cajal Grants RYC-2017-22482 and RYC-2020-030503-I, UPV/EHU PhD Grant 20/276, as well as from the EU FET-Open projects Quromorphic (828826) and EPIQUS (899368), Junta de Andaluc\'ia (P20-00617 and US-1380840), Valencian Government with reference number CIAICO/2021/184, Spanish Ministry of Economic Affairs and Digital Transformation through the QUANTUM ENIA project call – Quantum Spain project, and the European Union through the Recovery, Transformation and Resilience Plan – NextGenerationEU within the framework of the Digital Spain 2025 Agenda.
\end{acknowledgements}

\section*{Author contributions}
Y. Ding performed the initial simulations of the protocol on the D-Wave quantum annealer, J. Gonzalez-Conde. detailed and improved the implementation of the algorithm, M. Sanz and E. Solano provided the original idea based on a theoretical proposal by R.Or\'us, the rest of authors have contributed in the different phases of the development of the research.

\section*{Competing interests}
On behalf of all authors, the corresponding author states that there is no conflict of interest. 
\\
\section*{Data availability}
The available data that support the findings of this study are available at \url{https://github.com/yongchengding/DWaveFinancialCrashPrediction}.

\end{document}